\documentclass[12pt,a4paper]{article}
\usepackage{array}
\usepackage{fancyhdr}
\usepackage{graphicx}
\usepackage[notlof,nottoc,notlot]{tocbibind}
\usepackage[onehalfspacing]{setspace}
\usepackage{float}
\usepackage{caption}
\usepackage{subcaption}
\usepackage{hyperref}
\usepackage[a4paper,left=2.5cm, right=2.5cm, top=3cm, bottom=2.5cm]{geometry}
\usepackage[utf8]{inputenc}
\usepackage[toc,page]{appendix}
\title{Higher-order corrections for the deflection of light around a massive object}
\begin{document} 
\author{Carlos Rodriguez , Carlos A. \,Mar\'in  $^{1}$}
\date{\small{Universidad San Francisco de Quito}}
\maketitle

\begin{center}
{\large{}$^{1}$ cmarin@usfq.edu.ec}
\par\end{center}{\large \par}

%%%%%%%ABSTRACT%%%%%%%%
\begin{abstract}{From the Schwarzschild metric we obtain the higher-order terms (up to 20-th order) for the deflection of light around a massive object using the Lindstedt-Poincar\'e method to solve the equation of motion of a photon around the stellar object. Additionally, we obtain diagonal Pad\'e approximants from the perturbation expansion, and we show how these are a better fit for the numerical data. Furthermore, we use these approximants in ray-tracing algorithms to model the bending of light around the massive object.}
\end{abstract}
 keywords: general relativity, light deflection, Einstein, black hole, Pad\'e. 
\maketitle 

%%%%%%%INPUT YOUR FILES%%%%%%%%
\section*{Introduction}
The General Theory of Relativity (GTR) is probably one of the most elegant theories ever performed. It  was put forth by Albert Einstein in its current form in a 1916 publication, which expanded on his previous work of 1915  \cite{Einstein1,Einstein2}. This is summarized  in 14 equations \cite{Ohanian,Steven}. The Einstein field equations (ten equations written in tensor notation)
\medskip{}
\begin{eqnarray}
G_{\mu \nu}  \equiv R_{\mu \nu} - \frac{1}{2} R g_{\mu \nu} = k T_{\mu \nu} + \lambda g_{\mu \nu}
\label{eq:einsteinequation}
\end{eqnarray}
and the geodesic equations (4 equations)
\begin{equation}
\frac{d^{2}x^{\mu}}{ds^{2}} + \Gamma^{\mu}_{\rho \sigma}(\frac{dx^{\rho}}{ds})
(\frac{dx^{\sigma}}{ds}) = 0 \label{eq:geodesic}
\end{equation}
In (\ref{eq:einsteinequation}) $G_{\mu \nu}$ is the Einstein' s Tensor, which describes the curvature of space-time, $R_{\mu \nu}$ is the Ricci tensor, and $R$ is the Ricci scalar (the trace of the Ricci tensor), $g_{\mu \nu}$ is the metric tensor that describes the deviation of the Pythagoras theorem in a curved space, $T_{\mu \nu}$ is the stress-energy tensor describing the content of matter and energy. $k = \frac{ 8 \pi G}{c^{4}}$, where $c$ is the speed of light in vacuum and $G$ is the gravitational constant. Finally, $\lambda$ is the cosmological constant introduced by Einstein in 1917 \cite{Weinberg1,Weinberg2,Hobson} that is a measure of the contribution to  the energy density of the universe due to vacuum fluctuations.
In equation (\ref{eq:geodesic}) $s$ is the arc length satisfying the relation $ds^{2} = g_{\mu \nu}dx^{\mu}dx^{\nu}$ and $\Gamma_{\rho \sigma}^{\mu}$ are the connection coefficients ( Christoffel symbols of the second kind). $x^{\mu }$ is the position four-vector of the particle. We use Greek letters as $\mu , \nu , \alpha, $etc for 0,1,2,3. We have adopted the Einstein summation convention in which we sum over repeated indices.
Einstein's equations ( \ref{eq:einsteinequation})  tells us that the curvature of a region of  space-time is determined by the distribution of mass-energy of the same and they can be derived from the Einstein-Hilbert action \cite{Wald,Carroll}:

\begin{equation}
S = \int_{\Re} d^{4}x\sqrt{-g}\left[R-2kL_{F} + 2\lambda\right]
\end{equation}
where $\Re$ represents a region of space-time, $L_{F}$ is the Lagrangian density due to the fields of matter and energy and $g$ is the determinant of the metric tensor.

One of the most relevant predictions of General Relativity is the gravitational deflection of light. It was demonstrated during the solar eclipse of 1919 by two british expeditions \cite{Hawking}. One of the expeditions was led by Arthur Eddington and was bound for the island of Pr\'incipe in East Africa. The other one was led by Andrew Crommelin in the region of Sobral in Brazil. The light deflection can be measured taking a photograph of a star near the limb of the Sun, and then comparing it with another picture of the same star when the sun is not in the visual field. The observations are not easy. At present, Very Long Baseline Interferometry (VLBI) is used to measure the gravitational deflection of radio waves by the sun from observations of extragalactic radio sources  \cite{Lebach}. The result is very close to the value predicted by General Relativity \cite{Hobson} , which is  $\Omega = \frac{4 G M_{\Theta}}{R_{\Theta}c^{2}} = 1.752$ seconds of arc ($M_{\Theta}$ and $R_{\Theta}$ represent the solar mass and radius, respectively).

In the literature we can find calculations to second order of the deflection of light by a spherically symmetric body using Schwarzschild coordinates \cite{Bodenner,Freeman,Richter,Epstein} In this paper using the Schwarzschild metric we obtain  higher order corrections (up to 20-th order) for the gravitational deflection of light around a massive object like a star or a black hole using the Lindstedt-Poincar\'e method to solve the equation of motion of a photon around the stellar body. Additionally, we obtain diagonal Pad\'e approximants from the perturbation expansion, and we show how these are a better fit for the numerical data. We also  use these approximants in ray-tracing algorithms to model the bending of light around the massive object.

\medskip{}
\section{Schwarzschild metric}

\medskip{}
For a spherical symmetric space-time with a mass $M$ in the center of the coordinate system, the invariant interval is \cite{Misner,Kenyon}:

\begin{eqnarray}
\left(ds\right)^{2} = \gamma\left(c dt\right)^{2} - \gamma^{-1}\left(dr\right)^{2} -r^{2}\left(d\Omega \right)^{2}  \label{eq:MetricSchwar}
\end{eqnarray}
where $\left( d \Omega \right)^{2} = \left(d \theta\right)^{2} +sin^{2}\theta \left(d \phi\right)^{2}$ , with coordinates  $x^{0}=ct$, $x^{1}=r$, $ x^{2}=\theta$ 
and  $x^{3}=\phi$. $\gamma = 1 - \frac{r_{s}}{r}$ where $ r_{s} = \frac{2GM}{c^{2}}$
is the Schwarzschild radius.

The corresponding  covariant metric tensor is given by 

\begin{eqnarray}
g_{\mu\nu}=\left[\begin{array}{cccc}
\gamma & 0 & 0 & 0\\
0 & -\gamma^{-1} & 0 & 0\\
0 & 0 & -r^{2} & 0\\
0 & 0 & 0 & -r^{2}\sin^{2}\theta\end{array}\right]
\end{eqnarray}

Equation (\ref{eq:MetricSchwar}) has two singularities. The first one is when $r = r_{s}$
(the Schwarzschild radius) which defines the horizon event of a black hole. This is a mathematical singularity that can be removed by a convenient coordinate transformation like the one  introduced by Eddington in 1924 or Finkelstein in1958
\cite{Kenyon}:

\begin{eqnarray}
\hat{t} = t \pm \frac{r_{s}}{c} ln\vert \frac{r}{r_{s}} -1 \vert
\end{eqnarray}

With this coordinate transformation the invariant interval reads:

\begin{eqnarray}
\left(ds\right)^{2} = c^{2}\left(1 - \frac{r_{s}}{r}\right)\left(d \hat{t} \right)^{2} - 
\left(1 + \frac{r_{s}}{r}\right) \left(dr\right)^{2} \mp 2c \left(\frac{r_{s}}{r}\right) d\hat{t} dr
- r^{2} \left(d \Omega\right)^{2}
\end{eqnarray}

The other singularity in $r=0$ is a mathematical singularity. For a radius $r < r_{s}$, all massless and massive test particles eventually reach the singularity at $r = 0$. Thus, neglecting quantum effects like Hawking radiation \cite{Hawking,Hoyng}, any particle (even photons) that falls beyond this Schwarzschild radius will not escape the black hole.

\section{Geodesic equation for a photon in a Schwarzschild metric}
The geodesic equation can be written in an alternative form using the Lagrangian
\begin{eqnarray}
L\left(x^{\mu}, \frac{dx^{\mu}}{d\sigma}\right) = - g_{\alpha \beta}\left(x^{\mu}\right)
\frac{dx^{\alpha}}{d\sigma}\frac{dx^{\beta}}{d\sigma}
\end{eqnarray}
where $\sigma$ is a parameter of the trajectory of the particle, which is usually taken to be the proper time, $\tau$, or an affine parameter for massless particles like a photon. The resulting geodesic equation is:

\begin{eqnarray}
\frac{du_{\mu}}{d\sigma} = \frac{1}{2} \left(\partial_{\mu} g_{\alpha \beta}\right) u^{\alpha} u^{\beta}   \label{eq:geodesic2}
\end{eqnarray}
where $u_{\mu} = \frac{dx_{\mu}}{d \sigma}$.

Consider a photon traveling in the equatorial plane ($\theta = \pi / 2$) around a massive object. For a photon, $d\tau =0$ and thus, we use an affine parameter, $\lambda$, to describe the trajectory instead of the proper time, $\tau$. 
For the coordinates $ct$ ($\mu = 0$) and  $\phi$ ($\mu =2$) the geodesic equation  (\ref{eq:geodesic2}) give us, respectively :

\begin{eqnarray}
\frac{d}{d \lambda} \left[ \gamma c^{2} \frac{dt}{d \lambda}\right] = 0
\end{eqnarray}
and
\begin{eqnarray}
\frac{d}{d \lambda}\left[r^{2} \frac{d \phi}{d \lambda} \right] = 0
\end{eqnarray}

Both of these equations define the following constants along the trajectory of the photon around the massive object:

\begin{eqnarray}
\gamma c^{2} \frac{dt}{d \lambda} = E'
\end{eqnarray}
and
\begin{eqnarray}
r^{2} \frac{d \phi}{d \lambda} = J
\end{eqnarray}

where $E'$ has units of energy per unit mass and $J$ of angular momentum per unit mass (when $\lambda$ has units of time).

The invariant interval for the Schwarzschild metric in the plane $\theta = \pi / 2$ is.
\begin{eqnarray}
\left(ds\right)^{2} = c^{2}\left(d \tau\right)^{2}= \gamma  c^{2} \left(dt\right)^{2} - \gamma^{-1} \left(dr\right)^{2} -r^{2}\left(d \phi\right)^{2} = 0.
\end{eqnarray}
Using  $\frac{dr}{d\lambda} = \frac{dr}{d\phi} \frac{d\phi}{d\lambda}$ the last equation can be written in the form
\begin{eqnarray}
\gamma c^{2} \left( \frac{dt}{d\lambda}\right)^{2}  - \gamma^{-1} \left(\frac{dr}{d \phi} \right)^{2}\left(\frac{d\phi}{d\lambda}\right)^{2} - r^{2} \left(\frac{d\phi}{d\lambda}\right)^{2} = 0.  \label{eq:Schphi}
\end{eqnarray}
Multiplying (\ref{eq:Schphi}) by $\gamma$, and inserting the definitions of $E'$ and $J$
we obtain:
\begin{eqnarray}
\frac{\left(E'\right)^{2}}{c^{2}} -  \frac{J^{2}}{r^{4}}\left(\frac{dr}{d \phi} \right)^{2}
- \frac{\gamma J^{2}}{r^{2}} = 0
\end{eqnarray}

This equation can be turned into an equation for $U(\phi) = \frac{1}{r(\phi)}$, noting that 
\begin{equation}
\frac{dU}{d\phi} = - \frac{1}{r^{2}} \frac{dr}{d\phi}
\end{equation}

so we arrive at the following equation for $U(\phi)$:

\begin{eqnarray}
\frac{\left(E'\right)^{2}}{c^{2}} -  J^{2} \left(\frac{dU}{d \phi} \right)^{2}
- J^{2} U^{2} \left(1 - r_{s} U\right) = 0   \label{eq:orbit1}
\end{eqnarray}

By taking the derivative of equation (\ref{eq:orbit1}) with respect to $\phi$, we get the following differential equation for $U \left(\phi\right)$

\begin{eqnarray}
\left(\frac{dU}{d\phi}\right) \left( 2\frac{d^2U}{d\phi^2} + 2U - 3 r_s U^2 \right)=0 
\label{eq:orbit2}
\end{eqnarray}

The differential equation in (\ref{eq:orbit2}) can be separated into two differential equations for $U(\phi)$. The first one  is the equation for a photon that travels directly into or out from the black hole:

\begin{equation}
\frac{dU}{d\phi} = 0
\label{eq:orbit3}
\end{equation}

the other differential equation, applicable for trajectories in which $U(\phi)$ is not constant with respect to $\phi$, is the following:

\begin{equation}
\frac{d^2U}{d\phi^2} + U = \frac{3}{2} r_s U^2
\label{eq:orbit4}
\end{equation}

This equation can also be written in the following way, using the definition of the Schwarzschild radius:

\begin{equation}
\frac{d^2U}{d \phi^2}+U = \frac{3 G M U^2}{c^2}
\label{eq:orbitfinal}
\end{equation}

This is the equation for the trajectory of a massless particle that travels around a black hole in the equatorial plane.

\section{Differential equation for the trajectory of a photon}

In the previous section, we obtained a differential equation for a photon traveling around a masive object like a star or a black hole  (see equation (\ref{eq:orbitfinal})). This equation has an exact constant solution, for the unstable circular orbit of a photon around the black hole:

\begin{equation}
r_c =  \frac{3 G M}{c^2}
\end{equation}

where $r_c$ is the radius of the so-called photon sphere \cite{Misner}. We note that the radius of the photon sphere can be expressed in terms of the Schwarzschild radius:

\begin{equation}
r_c =  \frac{3 r_s}{2}
\end{equation}

The orbit described by a photon in the photon sphere is actually an unstable orbit , and a small perturbation in the orbit can lead either to the photon escaping the black hole or diving towards the event horizon \cite{Misner}.

Equation (\ref{eq:orbitfinal}) is nonlinear, and is highly difficult to solve analytically. However, a perturbative solution of this equation can be readily obtained. Let's first rewrite Equation (\ref{eq:orbitfinal}) in terms of $r_c$:

\begin{equation}
\label{eq:deflection-2}
\frac{d^2U}{d \phi^2}+U = r_c U^2
\end{equation}

\begin{figure}[ht]
\centering
\includegraphics[width=0.7\textwidth]{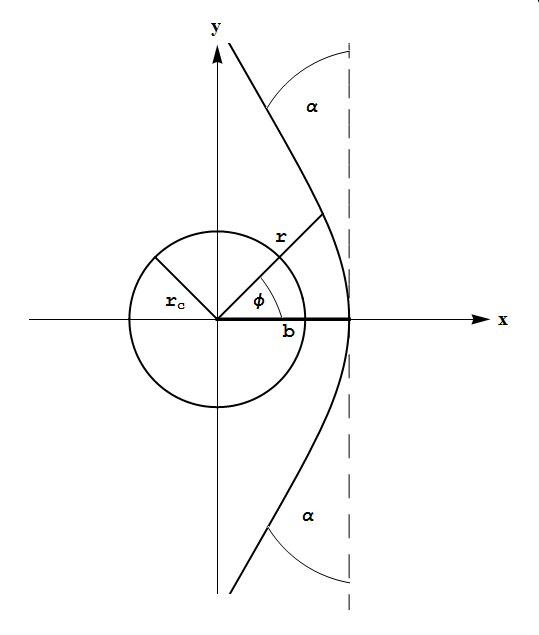}
\caption[Trajectory of a photon outside the photon sphere.]{Trajectory of a photon outside the photon sphere. The initial conditions are taken such that $r|_{\phi=0}=b$, and is called the impact parameter of the trajectory - the closest distance from the trajectory to the center of the black hole. We thus have, $\frac{dr}{d\phi}|_{\phi=0}=0$ and $\frac{dU}{d\phi}|_{\phi=0}=0 $. The photon experiences a total angular deflection of $2\alpha$. }
\label{fig:def01}
\end{figure}

Consider the initial conditions shown in Figure \ref{fig:def01}. The smallest value of the $r-$coordinate in the trajectory, $r=b$, is taken such that the photon escapes the black hole, $b>r_c$. We will rewrite Equation (\ref{eq:deflection-2}) in terms of $\epsilon = \frac{r_c}{b} < 1$, which we will use as a non-dimensional small number for our following perturbative expansions. Note that by multiplying both sides of Equation (\ref{eq:deflection-2}) by $b$, and defining the non-dimensional trajectory parameter

\begin{equation}
V (\phi) = \frac{b}{r (\phi)}  
\label{eq:V}
\end{equation} 

Equation (\ref{eq:deflection-2}), with the inclusion of the term $\epsilon = \frac{r_{c}}{b}$, then becomes a differential equation  in $V(\phi)$:

\begin{equation}
\label{eq:deflection-3}
\frac{d^2 V}{d\phi^2} +V = \epsilon V^2
\end{equation} 

where $0 < \epsilon = \frac{r_{c}}{b} < 1$, and with initial conditions given by

\begin{equation}
\label{eq:init-conditions-v}
V(\phi=0)=1\,;\,\frac{dV}{d\phi}(\phi=0)=0  
\end{equation}

Under these conditions, $V(\phi)$ is bounded such that

\begin{equation}
|V(\phi)|  \leq  1
\end{equation}

\section{First-order solution for $V(\phi)$}
A first idea to obtain a solution of Equation (\ref{eq:deflection-3}) is to consider a $V(\phi)$ as a power series in $\epsilon$:

\begin{equation}
\label{eq:deflection-4}
V(\phi;\epsilon)= V_0 (\phi) + \epsilon V_1(\phi) +\epsilon^2 V_2(\phi) + ...
\end{equation}

Plugging the expansion (\ref{eq:deflection-4}) into Equation (\ref{eq:deflection-3}) results in the following:

\begin{equation}
\label{eq:deflection-5}
\left(\frac{d^2 V_0}{d\phi^2} + \epsilon \frac{d^2 V_1}{d\phi^2} +  \epsilon^2 \frac{d^2 V_2}{d\phi^2} + ...\right) +(V_0 + \epsilon V_1 + \epsilon^2 V_2+...) = \epsilon \left(V_0 + \epsilon V_1 + \epsilon^2 V_2+...\right)^2
\end{equation}

We can group the powers of $\epsilon$ in Equation (\ref{eq:deflection-5}):

\begin{eqnarray}
  \epsilon^0 :&   \frac{d^2 V_0}{d\phi^2} + V_0 = 0      \label{eq:def6a}  
\end{eqnarray}
\begin{eqnarray}
  \epsilon^1 :&   \frac{d^2 V_1}{d\phi^2} + V_1 = {V_0}^2       \label{eq:def6b}     \end{eqnarray}
\begin{eqnarray} 
  \epsilon^2 :&   \frac{d^2 V_2}{d\phi^2} + V_2 = 2 V_0 V_1         \label{eq:def6c} \end{eqnarray}
\begin{eqnarray} 
  \epsilon^3 :&   \frac{d^2 V_3}{d\phi^2} + V_3 = (V_1)^2 +  2 V_0 V_2         \label{eq:def6d}
  \end{eqnarray}
  \begin{eqnarray}
  &\vdots   \nonumber
\end{eqnarray}

Note that the initial conditions of $V(\phi)$, applied to the asymptotic expansion in Equation (\ref{eq:deflection-4}), imply the following, by grouping powers of $\epsilon$:

\begin{eqnarray}
  \epsilon^0 :&   V_0(0)=1 \,;\, \frac{dV_0}{d\phi}(0)=0      \label{eq:inita} \end{eqnarray}
  \begin{eqnarray}
  \epsilon^k :&   V_k(0)=0 \,;\, \frac{dV_k}{d\phi}(0)=0 \,;\, k\geq1 \label{eq:initb} 
\end{eqnarray}

From these differential equations and initial conditions, we can readily obtain $V_0$ and $V_1$ iteratively \footnote{It is convenient to write the $V_k(\phi)$ in terms of polynomials in $cos(\phi)$}:

\begin{equation}
V_0(\phi)= cos (\phi)
\end{equation}

\begin{equation}
V_1 (\phi)= \frac{2}{3}-\frac{1}{3}cos (\phi)-\frac{1}{3} cos^2(\phi)
\end{equation}

Thus, we obtain an equation for $V(\phi)$, per Equation (\ref{eq:deflection-4}):

\begin{equation}
V(\phi) = cos (\phi) + \epsilon \left[\frac{2}{3}-\frac{1}{3}cos (\phi)-\frac{1}{3} cos^2(\phi) \right] + O(\epsilon^2)
\end{equation}

According to the coordinate system shown in Figure 1, the photon goes through a total angular deflection of $2\alpha$. This corresponds to setting $V(\phi)=0$ for both $\phi=\pi/2+\alpha$ and $\phi=-\pi/2-\alpha$. From both of these conditions considering that $\alpha$ is very small, to first order in $\epsilon$ we get: 
\begin{eqnarray}
\alpha = \frac{\frac{2 \epsilon}{3}}{\left(1 - \frac{\epsilon}{3}\right)} \label{eq:alpha1}
\end{eqnarray}
The total deviation of the photon is then
\begin{equation}
\Omega = 2\alpha  \approx  \frac{4 \epsilon}{3} = \frac{4 r_{c}}{3b} = \frac{4GM}{bc^{2}}
\label{eq:alpha2}
\end{equation}

For a light ray grazing the Sun' s  limb $b = R_{\Theta}  = 695510 km$ \cite{Olive} and we get the very well known value
\begin{equation}
\Omega = 2\alpha  \approx   \frac{4GM_{\Theta}}{R_{\Theta}c^{2}} = 1.7516  \quad arc seconds
\end{equation}
where $M_{\Theta} = 1.9885 \times 10^{30} kg$ is the Sun' s Mass, and $c = 2.99792458 \times 10^{6} \left[\frac{m}{s}\right]$ is the value of the speed of light in vacuum \cite{Olive}.

\section{Towards a second-order solution for $\Omega(\epsilon)$}

We will now see how to obtain higher-order solutions for $\Omega$. The differential equation in (\ref{eq:def6c}) has the following solution:

\begin{eqnarray}
\label{eq:wrong-second-order}
V_2(\phi) = -\frac{4}{9}+\frac{41}{36} cos(\phi) + \frac{2}{9} cos^2(\phi)+\frac{1}{12} cos^3(\phi) + \frac{5}{12} \phi sin(\phi)
\end{eqnarray}

However, the term in Equation (\ref{eq:wrong-second-order}) that goes as $\phi sin(\phi)$ grows without bound, and occurs because the right-handed side of Equation (\ref{eq:def6c}) contains terms proportional to the homogeneous solution of Equation (\ref{eq:def6c}): $a \, cos (\phi) + b \, \sin (\phi)$. When this happens, the solution contains terms that grow without bound, such as $\phi sin(\phi)$, called \textit{secular terms} \cite{Bush}. Thus, if we naively include Equation (\ref{eq:wrong-second-order}) in $V(\phi)$, our solution is no longer bounded. Thus, we have to eliminate any and all secular term that arises to arrive at a well-behaved solution for $V(\phi)$. \\
One method to do this, due to Lindstedt and Poincar\'e, is by solving the differential equation in the following \textit{strained coordinate} \cite{Bush}:

\begin{equation}
\label{eq:strained}
\tilde{\phi}= \phi \left( 1 + \omega_1 \epsilon + \omega_2 \epsilon^2 + \ldots \right)
\end{equation}

Where the $\omega_k$ are constants to be determined. In terms of this new strained coordinate $\tilde{\phi}$, Equation (\ref{eq:deflection-3}) becomes

\begin{equation}
\label{eq:dif-V-tilde}
\left(1 + \omega_1 \epsilon + \omega_2 \epsilon^2 + \ldots \right)^2 \frac{d^2V}{d\tilde{\phi}^2} + V(\tilde{\phi}) = \epsilon V^2(\tilde{\phi})
\end{equation}

We proceed in the previous way, and assume an asymptotic expansion on $V(\tilde{\phi})$:

\begin{equation}
\label{eq:expansion-V-tilde}
V(\tilde{\phi};\epsilon)= V_0 (\tilde{\phi}) + \epsilon V_1(\tilde{\phi}) +\epsilon^2 V_2(\tilde{\phi}) + ...
\end{equation}  \\

Plugging  the expansion(\ref{eq:expansion-V-tilde}) in Equation (\ref{eq:dif-V-tilde}), we obtain:

\begin{eqnarray}
\left(1 + \omega_1 \epsilon + \omega_2 \epsilon^2 + \ldots \right)^2 \left(\frac{d^2 V_0}{d\tilde{\phi}^2}   + \epsilon \frac{d^2 V_1}{d\tilde{\phi}^2} +  \epsilon^2 \frac{d^2 V_2}{d\tilde{\phi}^2} + ...\right)  + \nonumber \\
+(V_0 + \epsilon V_1 + \epsilon^2 V_2+...)  
  = \epsilon \left(V_0 + \epsilon V_1 + \epsilon^2 V_2+...\right)^2  \label{eq:V-tilde-expansion}
\end{eqnarray}
We can group the powers of $\epsilon$ in Equation (\ref{eq:V-tilde-expansion}):

\begin{eqnarray}
  \epsilon^0 :&   \frac{d^2 V_0}{d\tilde{\phi}^2} + V_0 = 0      \label{eq:Vdif0} \end{eqnarray}
 \begin{eqnarray}
  \epsilon^1 :&   \frac{d^2 V_1}{d\tilde{\phi}^2} + V_1 = {V_0}^2 - 2 \omega_1  \frac{d^2 V_0}{d\tilde{\phi}^2}       \label{eq:Vdif1}
  \end{eqnarray}
  \begin{eqnarray}
  \epsilon^2 :&   \frac{d^2 V_2}{d\tilde{\phi}^2} + V_2 = 2 V_0 V_1 - ({\omega_1}^2+2\omega_2 ) \frac{d^2 V_0}{d\tilde{\phi}^2} - 2 \omega_1 \frac{d^2 V_1}{d \tilde{\phi}^2}        \label{eq:Vdif2} 
  \end{eqnarray}
  \begin{eqnarray}
  \epsilon^3 :&   \frac{d^2 V_3}{d\tilde{\phi}^2} + V_3 = {V_1}^2 +  2 V_0 V_2 - (2 \omega_1 \omega_2 + 2 \omega_3) \frac{d^2 V_0}{d\tilde{\phi}^2} - ( {\omega_1}^2+2\omega_2)\frac{d^2 V_1}{d\tilde{\phi}^2} - 2 \omega_1 \frac{d^2 V_2}{d\tilde{\phi}^2}       \label{eq:Vdif3} 
  \end{eqnarray}
  \begin{eqnarray}
  &\vdots   \nonumber
  \end{eqnarray}
  With some care due to the definitions of the scaled variable and its derivative, we arrive at initial conditions for the $V_k(\tilde{\phi})$ from the initial conditions of $V(\phi)$:

\begin{eqnarray}
  \epsilon^0 :&   V_0( \tilde{\phi}=0)=1 \,;\, \frac{dV_0}{d\tilde{\phi}}(0)=0      \label{eq:init-tilde-a}
 \end{eqnarray}
 \begin{eqnarray}
  \epsilon^k :&   V_k(\tilde{\phi}=0)=0 \,;\, \frac{dV_k}{d\tilde{\phi}}(0)=0 \,;\, k\geq1 \label{eq:init-tilde-b}\
\end{eqnarray}

Solving the differential equation (\ref{eq:Vdif0}) with initial conditions (\ref{eq:init-tilde-a}), we arrive at the zeroth-order  contribution to $V(\tilde{\phi})$:

\begin{equation}
V_0 (\tilde{\phi}) = cos(\tilde{\phi})
\end{equation}

Similarly, we can obtain $V_1(\tilde{\phi})$ from Equation (\ref{eq:Vdif1}) subject to initial conditions (\ref{eq:init-tilde-b}):

\begin{equation}
V_1 (\tilde{\phi}) = \frac{2}{3} - \frac{1}{3} cos(\tilde{\phi}) -\frac{1}{3} cos^2 (\tilde{\phi}) + \omega_1 \tilde{\phi} sin(\tilde{\phi})
\end{equation}

We note that a secular term has appeared for $V_1(\tilde{\phi})$. However, we use our freedom in the definition of $\omega_1$ to eliminate this secular term by setting 

\begin{equation}
\omega_1 = 0
\end{equation}

so that the final form of $V_1(\tilde{\phi})$ is:

\begin{equation}
V_1 (\tilde{\phi}) = \frac{2}{3} - \frac{1}{3} cos(\tilde{\phi}) -\frac{1}{3} cos^2 (\tilde{\phi})
\end{equation}

Similarly we obtain for   $V_2(\tilde{\phi})$:

\begin{equation}
V_2 (\tilde{\phi}) = -\frac{4}{9} + \frac{5}{36} cos(\tilde{\phi}) +\frac{2}{9} cos^2 (\tilde{\phi}) +\frac{1}{12} cos^3 (\tilde{\phi})+ \frac{1}{144}(144 \omega_2+ 60) \tilde{\phi} sin(\tilde{\phi})
\end{equation}

To eliminate the secular term in $V_2 (\tilde{\phi})$, we set 

\begin{equation}
\omega_2 = -\frac{5}{12}
\end{equation}

and obtain the well-behaved second-order term

\begin{equation}
V_2 (\tilde{\phi}) = -\frac{4}{9} + \frac{5}{36} cos(\tilde{\phi}) +\frac{2}{9} cos^2 (\tilde{\phi}) +\frac{1}{12} cos^3 (\tilde{\phi})
\end{equation}
From all the solutions obtained so far, we can obtain the second-order correction to $\Omega(\epsilon).$ Note that $V(\tilde{\phi})$ is given by:

\begin{eqnarray}
V(\tilde{\phi})=cos(\tilde{\phi})+ \epsilon\left( \frac{2}{3} - \frac{1}{3} cos(\tilde{\phi}) -\frac{1}{3} cos^2 (\tilde{\phi}) \right) \nonumber \\
+ \epsilon^2 \left( -\frac{4}{9} + \frac{5}{36} cos(\tilde{\phi}) +\frac{2}{9} cos^2 (\tilde{\phi}) 
 +\frac{1}{12} cos^3 (\tilde{\phi})+ \right) + O(\epsilon^3) \label{eq:v-2-cos} 
\end{eqnarray}

We set up $\tilde{\phi}=\pi/2+\tilde{\alpha}$ in Equation (\ref{eq:v-2-cos}), such that $V(\pi/2+\tilde{\alpha}) = 0$ and obtain:

\begin{eqnarray}
-sin(\tilde{\alpha})+ \epsilon\left( \frac{2}{3} + \frac{1}{3} sin(\tilde{\alpha}) -\frac{1}{3} sin^2 (\tilde{\alpha}) \right) \nonumber \\
+ \epsilon^2 \left( -\frac{4}{9} - \frac{5}{36} sin(\tilde{\alpha}) +\frac{2}{9} sin^2 (\tilde{\alpha})  -\frac{1}{12} sin^3 (\tilde{\alpha})+ \right)  + O(\epsilon^3) = 0 \label{eq:v-2-sin}
\end{eqnarray}

We could truncate this Equation and solve the resultant cubic polynomial in $sin(\tilde{\alpha})$. However, this method would not be easy to generalize, because we do not have a general formula for the roots of fifth-order polynomials and above, according to Galois theory \cite{Tignol}. Also, an $n-$th order polynomial results in $n$ different complex solutions, one of which we expect to have a leading term of order $\epsilon$, to obtain a better approximation of $\Omega$, and we would need to check all the $n$ different solutions for this. Additionally, we have to remember that so far this is an asymptotic expansion in $\epsilon$, and the truncation of the higher-order terms does not allow us to clearly see what the order of our estimate for $\Omega(\epsilon)$ is. All of these problems are solved by assuming that $sin(\tilde{\alpha})$ has the following expansion in $\epsilon$, with a leading term of order $\epsilon^1$:

\begin{equation}
sin(\tilde{\alpha}) = \epsilon \chi_1 + \epsilon^2 \chi_2 + \epsilon^3 \chi_3 + \ldots
\end{equation}

where the $\chi_k$ are constants to be determined. Inserting this new expansion into Equation (\ref{eq:v-2-sin}) leads to the following algebraic Equation:

\begin{equation}
\left(\frac{2}{3}-\chi_1 \right) \epsilon + \left( -\frac{4}{9} +\frac{\chi_1}{3} - \chi_2 \right) \epsilon^2 + O(\epsilon^3) = 0
\end{equation}

Then, we have to equal to zero the different powers of $\epsilon$ in the last equation. Equating to zero the terms with $\epsilon^1$  we arrive at:

\begin{equation}
\chi_1=\frac{2}{3}
\end{equation}

and equating to zero the terms with $\epsilon^2$ we arrive at:

\begin{equation}
\chi_2=-\frac{2}{9}
\end{equation}

Thus, $sin(\tilde\alpha)$ is given by:

\begin{equation}
sin(\tilde\alpha) = \frac{2}{3} \epsilon  - \frac{2}{9} \epsilon^2 + O(\epsilon^3)
\end{equation}

To obtain $\tilde\alpha$, we employ the Taylor series of $arcsin(x)$ around $x=0$:

\begin{equation}
arcsin(x) = x +\frac{x^3}{6} + O(x^5)
\end{equation}

and obtain

\begin{equation}
\tilde\alpha = \frac{2}{3} \epsilon  - \frac{2}{9} \epsilon^2 + O(\epsilon^3)
\end{equation}

However, what we actually want is $\alpha$. From the definition of the strained coordinate $\tilde\phi$ in (\ref{eq:strained}), it is clear that:

\begin{equation}
\frac{\pi}{2} + \alpha=\frac{\frac{\pi}{2} + \tilde\alpha}{1+\omega_1 \epsilon + \omega_2 \epsilon^2 + O(\epsilon^3)}  
\end{equation}

From the last  equation, an using the Taylor expansion of $\frac{1}{1+x} = \sum_{n = 0}^{\infty} (-1)^{n}x^{n}$ around $x=0$, we obtain:

\begin{equation}
\alpha = \frac{2 }{3}\epsilon  + \left( \frac{5 \pi}{24} -\frac{2}{9} \right) \epsilon^2 + O(\epsilon^3)
\end{equation}

From which we can obtain the total deflection angle, $\Omega = 2 \alpha$

\begin{equation}
\Omega = \frac{4 }{3}\epsilon  + \left( \frac{5 \pi}{12} -\frac{4}{9} \right) \epsilon^2 + O(\epsilon^3) = \frac{4 GM}{b c^2} + \left( \frac{15 \pi}{4} - 4 \right) \left( \frac{GM}{b c^2}\right)^2 
+ O\left[ \left( \frac{GM}{b c^2}\right)^3 \right]
\end{equation}
This result is in agreement with other work \cite{Bodenner,Freeman,Richter,Epstein}.

\section{Higher order solutions for $\Omega (\epsilon)$}
The previous procedure can be automated to obtain higher-order expressions for $\Omega$. Notably, all the solutions for the $V_k(\tilde\phi)$ are in the forms of $(k+1)-$order polynomials of $cos(\tilde\phi)$, and a secular term that is eliminated by choosing a suitable $\omega_k$. The use of the expansion of $\sin(\tilde\alpha)$ in powers of $\epsilon$ guarantees both the form of $sin(\tilde\alpha)$ with a leading term of order $\epsilon$, and leads to algebraic equations for the $\chi_k$ that are exceedingly easy to solve. Notably, getting a higher-order solution conserves the lower-order terms. Consider the formal Taylor expansion of $\Omega$ around $\epsilon=0$:

\begin{equation}
\label{eq:omega-kappa}
 \Omega = \kappa_1 \epsilon^1 + \kappa_2 \epsilon^2 + \kappa_3 \epsilon^3 +\ldots
\end{equation}

A table of the coefficients $\kappa_n$ of the series of $\Omega$ in (\ref{eq:omega-kappa}) can be found in Table \ref{tab-kappa}. These $\kappa_n$ were found using the method of the previous sections, and obtaining the solutions up to $V_{20}(\tilde\phi)$.

\begin{center}
\begin{tabular}{|c|c|c|}
\hline 
& Exact value & Numerical value \\
\hline 
$\kappa_1$ & $\frac{4}{3}$ & 1.33333\\
\hline
$\kappa_2$ & $\frac{5 \pi}{12} - \frac{4}{9}$ & 0.864552\\
\hline
$\kappa_3$ & $\frac{122}{81} - \frac{5\pi}{18}$ & 0.633508\\
\hline
$\kappa_4$ & $\frac{385 \pi}{576} - \frac{130}{81}$ &0.494911\\
\hline
$\kappa_5$ & $\frac{7783}{2430} - \frac{385 \pi}{432}$ & 0.403082\\
\hline
$\kappa_6$ & $\frac{103565 \pi}{62208} - \frac{21397}{4374}$ & 0.338319\\
\hline
$\kappa_7$ & $ \frac{544045}{61236}-\frac{85085\pi}{31104} $ &  0.290571\\
\hline
$\kappa_8$ & $\frac{6551545  \pi}{1327104} - \frac{133451}{8748}$ & 0.254143\\
\hline
$\kappa_9$ & $\frac{1094345069}{39680928} - \frac{116991875 \pi}{13436928}$ & 0.225577 \\
\hline
$\kappa_{10}$ & $\frac{2268110845 \pi}{143327232} - \frac{1091492587}{22044960}$ & 0.202655\\
\hline
$\kappa_{11}$ & $\frac{33880841953}{374134464} - \frac{18553890355 \pi}{644972544}$ & 0.183902 \\
\hline
$\kappa_{12}$ & $\frac{3278312542505 \pi}{61917364224} - \frac{627972527}{3779136}$ & 0.168300 \\
\hline
$\kappa_{13}$ & $\frac{17954674772417}{58364976384} - \frac{1514986498025 \pi}{15479341056}$ & 0.155132\\
\hline
$\kappa_{14}$ & $\frac{135335969751125 \pi}{743008370688} - \frac{53937207017735}{94281884928}$ & 0.143875\\
\hline
$\kappa_{15}$ & $\frac{1532445398265737}{1432594874880} - \frac{1138317723327785 \pi}{3343537668096}$ & 0.134145 \\
\hline
$\kappa_{16}$ & $\frac{1094325341294717675  \pi}{1711891286065152} - \frac{4027582104301883}{2005632824832}$ & 0.125654 \\
\hline
$\kappa_{17}$ & $\frac{2064610875963794827}{545532128354304} - \frac{128887453213429625 \pi}{106993205379072}$ &0.118179\\
\hline
$\kappa_{18}$ & $\frac{1263396148548501892925 \pi}{554652776685109248} - \frac{2657173119021192719}{371328591568896}$ & 0.111548\\
\hline
$\kappa_{19}$ & $\frac{1085138496158025821251}{79959423384502272} - \frac{399330245672667033725 \pi}{92442129447518208}$ & 0.105625\\
\hline
$\kappa_{20}$ & $\frac{218695963585074038928865 \pi}{26623333280885243904} - \frac{75186822805298075761}{2913501256925184}$ & 0.100303\\
\hline

\end{tabular}
\captionof{table}[Coefficients $\kappa_n$ of the series of $\Omega$ in (\ref{eq:omega-kappa}).]{Coefficients $\kappa_n$ of the series of $\Omega$ in (\ref{eq:omega-kappa}). For these coefficients, we report both the exact values and the numerical values with 6 significant figures.}
\label{tab-kappa}
\end{center}

Clearly, as $\epsilon \rightarrow 1$, $\Omega(\epsilon) \rightarrow \infty$, because the photon starts going around the black hole as it starts closing in the photon sphere ($b \rightarrow r_c$). This means that $\Omega (\epsilon)$ has a singularity at $\epsilon=1$. The Taylor expansion of $\Omega(\epsilon)$ around $\epsilon = 0$ that we found at Equation (\ref{eq:omega-kappa}) does not return an estimate for the position of this singularity, because a polynomial does not have a singularity. However, we can obtain Pad\'e approximants for $\Omega$ around $\epsilon = 0$, and these will return an estimate for the position of this singularity.

As a small refresher on Pad\'e approximants, we note their definition. A Pad\'e approximant of a function $f(x)$ is a rational function $f^{[L|M]}(x)$  of the form:

\begin{equation}
f^{[L|M]} (x) = \frac{a_0 + a_1 x + a_2 x^2 + ... + a_L x^L}{1+b_1 x + b_2 x^2 + ... +b_M x^M}
\end{equation}

where $f(x)$ and $f^{[L|M]}(x)$ are equal in their first $L+M+1$ derivatives around $x=0$ \cite{Saff,Yamada}. A diagonal Pad\'e approximant $f^{[N]}(x)$ is a Pad\'e approximant in which $N=L=M$. We can obtain the diagonal Pad\'e approximants for up to N=10 with the Taylor series expansion for $\Omega(\epsilon)$. For example, the $\Omega^{[1]}(\epsilon)$ Pad\'e approximant is given by:

\begin{equation}
\Omega^{[1]}(\epsilon)= \frac{48 \pi + \left(64 + 16 \pi -15 \pi^2 \right)\epsilon}{96 + (32-30 \pi) \epsilon}
\end{equation}

The exact formulas for the Pad\'e approximants of $\Omega(\epsilon)$ are rather complicated because of the powers of $\pi$ involved. Due to this, our work with Pad\'e approximants will be purely numeric. All the  Pad\'e approximants $\Omega^{[N]}(\epsilon)$ have a singularity of order 1 at a position around $\epsilon=1$. The position of this singularity, $\epsilon_s$, is tabulated for the 10 Pad\'e approximants in  Table \ref{tab-singularities}.

\begin{center}
\begin{tabular}{|c|c|c|}
\hline 
 $N$ & $\epsilon_s$\\
\hline
1 & 1.54222\\
\hline
2 & 1.21736 \\
\hline
3 & 1.11036\\
\hline
4 & 1.06664\\
\hline
5 & 1.04532\\
\hline
6 & 1.03238\\
\hline
7 & 1.0245\\
\hline
8 & 1.01915\\
\hline
9 & 1.01537\\
\hline
10 & 1.01264\\
\hline
\end{tabular}
\captionof{table}{The position of the singularity near $\epsilon=1$ for the Pad\'e approximants, $\Omega^{[N]}(\epsilon)$.}
\label{tab-singularities}
\end{center}

\section{Numerical tests for $\Omega(\lambda)$ and its Pad\'e approximants}

\begin{figure}[h]
\centering
\includegraphics[width=0.9\textwidth]{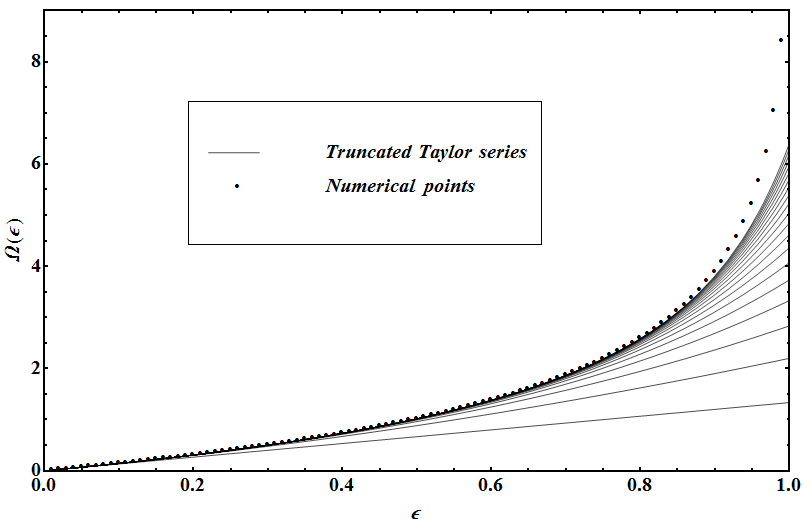}
\caption[Numerical points obtained for $\Omega(\epsilon)$ compared to the truncated $n-$th order Taylor polynomials of $\Omega(\epsilon)$.]{Numerical points obtained for $\Omega(\epsilon)$ compared to the truncated $n-$th order Taylor polynomials of $\Omega(\epsilon)$, up to $20-$th order. With increasing value of $n$, the polynomials take larger values.}
\label{fig:Taylor}
\end{figure}

\begin{figure}[h]
\centering
\includegraphics[width=0.9\textwidth]{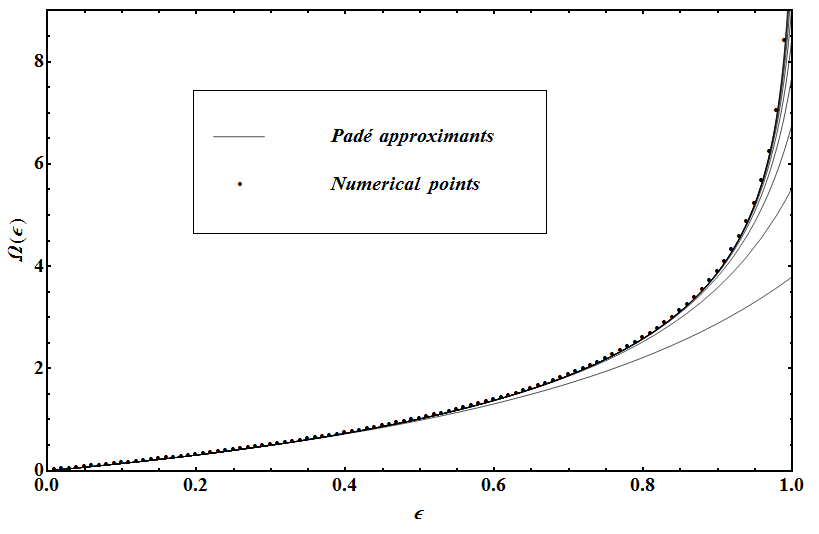}
\caption[Numerical points obtained for $\Omega(\epsilon)$ compared to the truncated $N-$th diagonal Pad\'e approximants of $\Omega(\epsilon)$.]{Numerical points obtained for $\Omega(\epsilon)$ compared to the truncated $N-$th diagonal Pad\'e approximants of $\Omega(\epsilon)$, up to $N=10$. With increasing value of $N$, the Pad\'e approximants take larger values. For $\epsilon=0.99$, the $N=10$ Pad\'e approximant is within 3\% of the numerical value of $\Omega(\epsilon)$.}
\label{fig:Pade}
\end{figure}

All the coefficients for the Taylor expansion of $\Omega(\epsilon)$ were obtained around $\epsilon=0$. We can test the correctness of the methods thus far used to obtain this function by comparing it to the results of numerical solutions of Equation (\ref{eq:orbitfinal}). This is done with both truncated $n-$th order Taylor polynomials from $\Omega(\epsilon)$ and for the Pad\'e approximants we obtain from this function, $\Omega^{[N]}(\epsilon)$. This comparisons are shown in Figures \ref{fig:Taylor} and \ref{fig:Pade}.

We see from Figures \ref{fig:Taylor} and \ref{fig:Pade} that the Pad\'e approximants are much faster at converging into the actual form of $\Omega(\epsilon)$. The convergence of the Pad\'e approximants is such that for $\epsilon=0.99$, the $N=10$ diagonal Pad\'e approximant is within 3\% of the corresponding numerical value. This is mainly due to the fact that Pad\'e approximants are better in approximating functions that have singularities \cite{Yamada}. Once we know that the $\Omega(\epsilon)$ behave correctly, we can use the $\Omega (\epsilon)$ to simulate the bending of light around a black hole. A simple first--order ray tracing algorithm that does this for the different approximations of $\Omega(\epsilon)$ we have found is shown in Appendix.
\
%%%%%%%%%%%%%%%%%%%%%%%%%%%%%%%%%%%%%%%%%%%%%%%%%%%%%%%%%%%%%%%%%%%%%%%%%%%%%%%%%%%%%%%%%%%%%%%%%%%%%%%%%%%%%
%%%%%%%%%%%%%%%%%%%%%%%%%%%%%%%%%%%%%%%%%%%%%%%%%%%%%%%%%%%%%%%%%%%%%%%%%%%%%%%%%%%%%%%%%%%%%%%%%%%%%%%%%%%%%

\section{Conclusions}
One of the most important predictions of the General Theory of Relativity is undoubtedly
the bending of light around a massive object like a star, a black hole or even a galaxy (in this case it can generate a gravitational lens) .  In this paper using the Schwarzschild metric we have obtained  higher order corrections for the gravitational deflection of light around  said objects using the Lindstedt-Poincar\'e method to solve the equation of motion of a photon around the stellar body. We have successfully obtained an expression for $\Omega\left(\epsilon\right)$ , the angular deflection experienced by a photon traveling around the massive object. We have assumed that the parameter $\epsilon$ was small, and we were able to obtain the coefficients $\kappa_n$ of the series of  $\Omega\left(\epsilon\right)$ up to  $V_{20}(\tilde\phi)$ (the non- dimensional trajectory parameter, see equation (\ref{eq:V})). The results are given in Table \ref{tab-kappa}.
Additionally, we have obtained diagonal Pad\'e approximants from the perturbation expansion, and we have  shown how these are a better fit for the numerical data. The best approximation for $\Omega\left(\epsilon\right)$ we obtained was consistent with the numerical data even for an $\epsilon \approx 0.99$. In this case, the $N = 10$ diagonal Pad\'e approximant is within $3\%$ of the corresponding numerical value.
 We were able to use  this estimate for $\Omega\left(\epsilon\right)$  in ray-tracing algorithms to model the bending of light around the massive object.
\clearpage
\appendix
\renewcommand{\theequation}{\Alph{section}.\arabic{equation}}
\section*{Appendix}
\section{Ray tracing using $\Omega(\epsilon)$}
% Reset equation numbering after each section from now on
\setcounter{equation}{0}

Consider an observer $A$ immersed in a background distribution of far away light sources. This observer can obtain the angular position of every object in the sky, and determine the intensity of light that comes from every point in the sky, $I_A(\theta,\phi)$, in spherical coordinates. Now, imagine another point in space, $B$, far enough from the observer $A$ such that the intensity of light that comes from every point in the sky, according to an observer in point $B$, is also given by the distribution found by observer $A$: $I_B(\theta,\phi)=I_A(\theta,\phi)$. If we place a black hole at point $B$, then light coming from the faraway sources will bend around the black hole such that the original observer will see a different distribution of light around the black hole. In this condition, the observer will be able to note that the black hole effectively subtends a solid angle in the sky -- region in the sky devoid of any light due to the black hole. One half of the angle subtended by the black hole will effectively give the ''angular radius'' of the black hole, as seen by the observer, $r_{BH}$. 

If we consider that the black region of the sky due to the black hole is due to the radius of the photon sphere, $r_c$, instead of the the Schwarzschild radius, $r_s$ \footnote{One can convince himself of this by considering the light from faraway objects that grazes the black hole at a distance given by $r=b$. This trajectory of this light is bended by the black hole for $b>r_c$. However, if $b<r_c$, the light will not escape and effectively no light coming from faraway objects will seem to originate from $r<r_c$, which becomes an effective radius for the black hole, according to observer $A$ in these conditions. In the case that mass enters the black hole, and emits light from an $r$ that obeys  $r_s<r<r_c$, the light \textit{can} escape the black hole, and is severely red-shifted. However, we are here considering a black hole with no light sources between $r_s<r<r_c$.}, and if we choose the coordinate system such that the black hole is at the positive x-axis, $\theta=\pi/2$ and $\phi=0$, then, by definition of $r_{BH}$, the new distribution of light measured by the observer will obey (for small enough $r_{BH}$):

\begin{equation}
I(\theta, \phi) =0 \, ; \,  (\theta-\pi/2)^2 + \phi^2 \leq (r_{BH})^2
\end{equation}

\begin{figure}[h]
\centering
\includegraphics[width=0.7\textwidth]{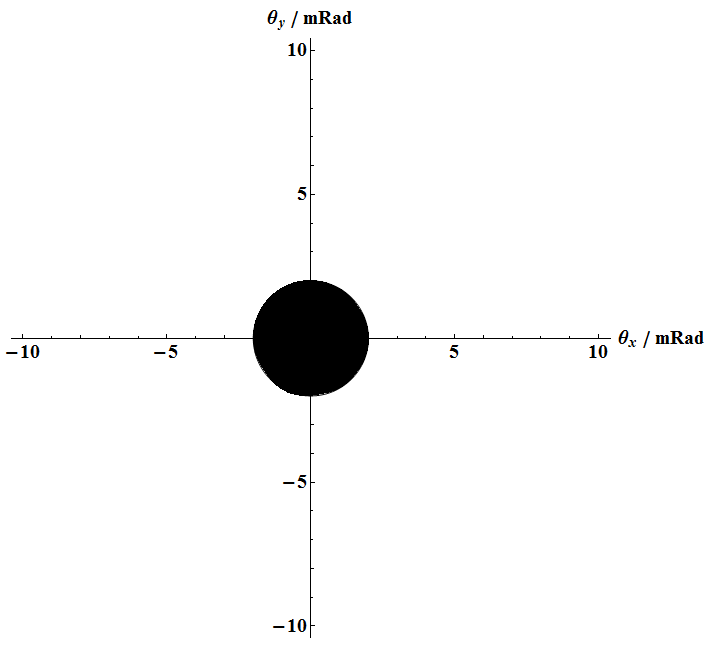}
\caption[A black hole with $r_{BH}=2$ mrad in the center of the $(\theta_x,\theta_y)$ coordinate system.]{A black hole with $r_{BH}=2$ mrad in the center of the $(\theta_x,\theta_y)$ coordinate system.}
\label{fig:BHFig1}
\end{figure}

For other values of $(\theta,\phi)$, the observer sees light distribution shifted by the  $\Omega(\epsilon)$, where $\epsilon$ is given by:

\begin{equation}
\epsilon = \frac{r_c}{b}  = \frac{r_{BH}}{\sqrt{(\theta-\pi/2)^2 + \phi^2}}
\end{equation}

for $\theta \approx \pi/2$. We can use a further simplification of this latter equation, and use the coordinates $(\theta_x,\theta_y)$ defined by $\theta_x = \phi$, $\theta_y = \theta - \pi/2$. For small values of  $\theta_x$ and $\theta_y$, say, in the order of milliradians, we can write:

\begin{equation}
I(\theta_x, \theta_y) =0 \, ; \,  \theta^2_x + \theta^2_y  \leq (r_{BH})^2
\end{equation}

and

\begin{equation}
\epsilon = \frac{r_{BH}}{\sqrt{\theta^2_x + \theta^2_y}}
\end{equation}

where the analogue with Cartesian coordinates is evident. This coordinate system is shown in Figure \ref{fig:BHFig1} for a black hole that subtends $4 \pi \times 10^{-6}$ steradians, such that $r_{BH} = 2$ mrad.

\begin{figure}[h]
\centering
\includegraphics[width=0.7\textwidth]{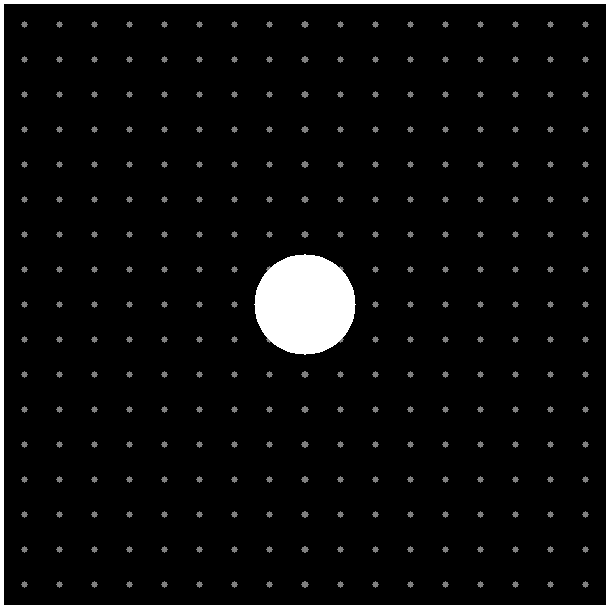}
\caption[Intensity due to background light sources, $I_A(\theta_x,\theta_y)$, without a black hole present.]{$600 \times 600 $ image corresponding to the intensity due to background light sources, $I_A(\theta_x,\theta_y)$, without a black hole present. Each pixel corresponds to 1 mrad. The big star, in white, has a radius of 50 mrad. The small stars, in gray, have a radius of 3 mrad. The star is at the center of the coordinate system, $(\theta_x,\theta_y)=(0,0)$.}
\label{fig:BHIA}
\end{figure}

This coordinate choice allows one to define the distribution of intensities that observer $A$ sees to be (disregarding some attenuation factors):

\begin{equation}
\label{eq:RayTracing}
I(\theta_x, \theta_y) =I_A \left( \theta_x - \Omega (\epsilon ) \frac{\theta_x}{\sqrt{\theta^2_x+\theta^2_y}}, \theta_y - \Omega (\epsilon ) \frac{\theta_y}{\sqrt{\theta^2_x+\theta^2_y}} \right)       \, ; \,  \theta^2_x + \theta^2_y  \leq ( r_{BH})^2
\end{equation}

where we have used $I_A(\theta_x,\theta_y)$, the angular distribution of intensities seen by observer $A$ without the black hole present, and using the coordinates $(\theta_x,\theta_y)$.  We can see the effect of applying equation (\ref{eq:RayTracing}) by using the $I_A(\theta_x,\theta_y)$ defined from Figure \ref{fig:BHIA}. 

\begin{figure}[h]
\centering
\includegraphics[width=0.7\textwidth]{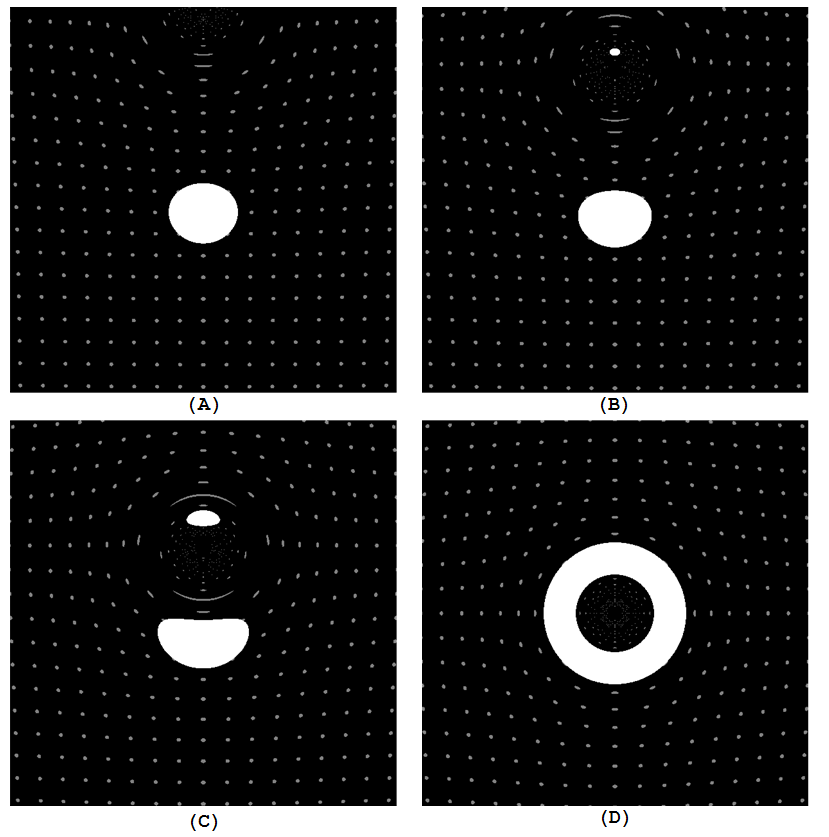}
\caption[Background of Figure \ref{fig:BHIA} warped by a black hole, employing the truncated first-order Taylor polynomial of $\Omega(\epsilon)$.]{Background of Figure \ref{fig:BHIA} warped by a black hole at $(A)$ $\theta_y = 300$ mrad, $(B)$ $\theta_y = 200$ mrad, $(C)$ $\theta_y = 100$ mrad, and $(D)$ $\theta_y = 0$ mrad. We make use of the truncated first-order Taylor polynomial of $\Omega(\epsilon)$.}
\label{fig-BH-Taylor-1}
\end{figure}

\begin{figure}[h]
\centering
\includegraphics[width=0.7\textwidth]{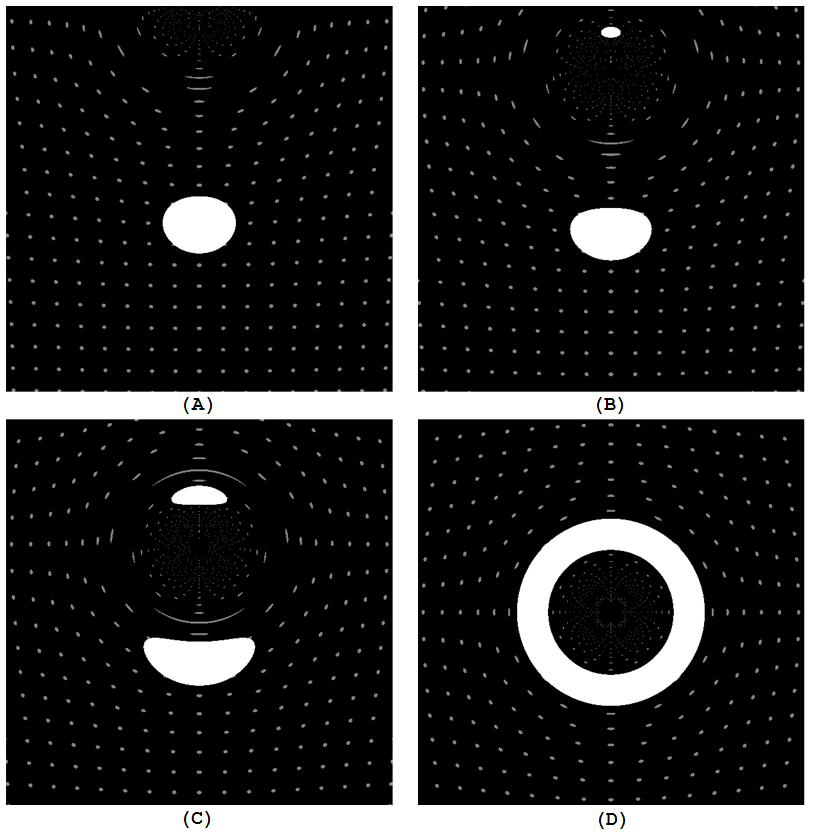}
\caption[Background of Figure \ref{fig:BHIA} warped by a black hole, employing the diagonal $N=2$ Pad\'e approximant of $\Omega(\epsilon)$.]{Background of Figure \ref{fig:BHIA} warped by a black hole at $(A)$ $\theta_y = 300$ mrad, $(B)$ $\theta_y = 200$ mrad, $(C)$ $\theta_y = 100$ mrad, and $(D)$ $\theta_y = 0$ mrad. We make use of the diagonal $N=2$ Pad\'e approximant of $\Omega(\epsilon)$.}
\label{fig-BH-Pade-2}
\end{figure}

\begin{figure}[h]
\centering
\includegraphics[width=0.7\textwidth]{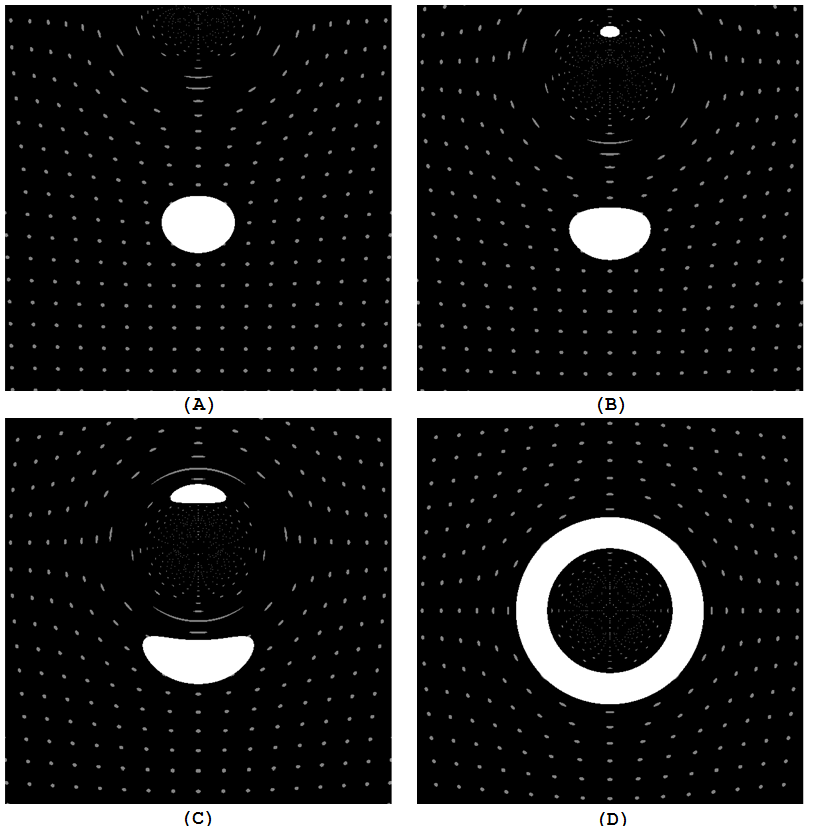}
\caption[Background of Figure \ref{fig:BHIA} warped by a black hole, employing the diagonal $N=10$ Pad\'e approximant of $\Omega(\epsilon)$.]{Background of Figure \ref{fig:BHIA} warped by a black hole at $(A)$ $\theta_y = 300$ mrad, $(B)$ $\theta_y = 200$ mrad, $(C)$ $\theta_y = 100$ mrad, and $(D)$ $\theta_y = 0$ mrad. We make use of the diagonal $N=10$ Pad\'e approximant of $\Omega(\epsilon)$.}
\label{fig-BH-Pade-10}
\end{figure}

To model the deflection of light with the distribution in Figure \ref{fig:BHIA}, we use Equation (\ref{eq:RayTracing}) with $\Omega(\epsilon)$ approximated as a truncated first-degree Taylor polynomial, and as diagonal Pad\'e approximants with $N=2$ and $N=10$. The resulting images can be found in Figures \ref{fig-BH-Taylor-1} to \ref{fig-BH-Pade-10}. We use a black hole with $r_{BH} = 10 $ mrads.

The most notable difference between Figures \ref{fig-BH-Taylor-1} and \ref{fig-BH-Pade-2} is the position of the white ring around the black hole, corresponding to the gravitational lensing of the big, white star at the black hole position $\theta_y=0$. When using a better approximation of $\Omega(\epsilon)$, this ring has greater inner and outer radii, and is thinner. In Figure \ref{fig-BH-Pade-10}, there are 8 white pixels around $\theta^2_x+\theta^2_y = 10$ millirads, corresponding to a second ring of light due to the big, white star.

\clearpage

%%%%%%%REFERENCES%%%%%%%%

\end{document}